\documentstyle[12pt]{article}
\font\tenrm=cmr10

\begin{document}
\renewenvironment{thebibliography}[1]
  { \begin{list}{\arabic{enumi}.}
    {\usecounter{enumi} \setlength{\parsep}{0pt}
     \setlength{\itemsep}{3pt} \settowidth{\labelwidth}{#1.}
     \sloppy
    }}{\end{list}}

\newcommand{\EQ}{\begin{equation}}
\newcommand{\EN}{\end{equation}}
\newcommand{\bea}{\begin{eqnarray}}
\newcommand{\ena}{\end{eqnarray}}
\newcommand{\vs}[1]{\vspace{#1 mm}}
\renewcommand{\a}{\alpha}
\renewcommand{\b}{\beta}
\renewcommand{\c}{\gamma}
\renewcommand{\d}{\delta}

\newcommand{\th}{\theta}
\newcommand{\Th}{\Theta}
\newcommand{\TH}{\Theta}
\newcommand{\pa}{\partial}
\newcommand{\g}{\gamma}
\newcommand{\G}{\Gamma}
\newcommand{\A}{\Alpha}
\newcommand{\B}{\Beta}
\newcommand{\D}{\Delta}
\newcommand{\e}{\epsilon}
\newcommand{\E}{\Epsilon}
\newcommand{\z}{\zeta}
\newcommand{\Z}{\Zeta}
\newcommand{\k}{\kappa}
\newcommand{\K}{\Kappa}
\renewcommand{\l}{\lambda}
\renewcommand{\L}{\Lambda}
\newcommand{\m}{\mu}
\newcommand{\M}{\Mu}
\newcommand{\n}{\nu}
\newcommand{\N}{\Nu}
\newcommand{\x}{\chi}
\newcommand{\X}{\Chi}
\newcommand{\p}{\pi}
\renewcommand{\P}{\Pi}
\newcommand{\r}{\rho}
\newcommand{\R}{\Rho}
\newcommand{\s}{\sigma}
\renewcommand{\S}{\Sigma}
\renewcommand{\t}{\tau}
\newcommand{\T}{\Tau}
\newcommand{\y}{\upsilon}
\newcommand{\Y}{\upsilon}
\renewcommand{\o}{\omega}
\renewcommand{\O}{\Omega}

\newcommand{\hs}[1]{\hspace{#1 mm}}
\newcommand{\shalf}{\frac{1}{2}}

\renewcommand{\Im}{{\rm Im}\,}
\newcommand{\NP}[1]{Nucl.\ Phys.\ {\bf #1}}
\newcommand{\PL}[1]{Phys.\ Lett.\ {\bf #1}}
\newcommand{\NC}[1]{Nuovo Cimento {\bf #1}}
\newcommand{\CMP}[1]{Comm.\ Math.\ Phys.\ {\bf #1}}
\newcommand{\PR}[1]{Phys.\ Rev.\ {\bf #1}}
\newcommand{\PRL}[1]{Phys.\ Rev.\ Lett.\ {\bf #1}}
\newcommand{\MPL}[1]{Mod.\ Phys.\ Lett.\ {\bf #1}}
\renewcommand{\thefootnote}{\fnsymbol{footnote}}

\newpage
\parindent=1.5pc
\begin{titlepage}
\begin{flushright}
IFUM 435/FT  \\
\end{flushright}
\vspace{2cm}
\begin{center}{{\bf QUANTUM SYMMETRIES IN \\
               \vglue 10pt
               SUPERSYMMETRIC TODA THEORIES\\}

\vglue 1.0cm
{SILVIA PENATI}\\
\baselineskip=14pt
{\it Dipartimento di Fisica dell'Universit\`a di Milano and }\\
\baselineskip=14pt
{\it INFN, Sezione di Milano,Via Celoria 16, I--20133 Milano, Italy}\\
\vglue 0.3cm
{and}\\
\vglue 0.3cm
{DANIELA ZANON}\\
\baselineskip=14pt
{\it Dipartimento di Fisica dell'Universit\`a di Milano and }\\
\baselineskip=14pt
{\it INFN, Sezione di Milano,Via Celoria 16, I--20133 Milano, Italy}\\
\vglue 0.8cm
{ABSTRACT}}
\end{center}
\vglue 0.3cm
{\rightskip=3pc
 \leftskip=3pc
 \tenrm\baselineskip=12pt
 \noindent
We consider two--dimensional supersymmetric Toda theories based on the
Lie superalgebras $A(n,n)$, $D(n+1,n)$ and $B(n,n)$ which admit a fermionic
set of simple roots and a fermionic untwisted affine extension. In
particular, we concentrate on two simple examples, the $B(1,1)$ and $A(1,1)$
theories.
Both in the conformal and  massive case
we address the issue of quantum integrability by
constructing the first non trivial conserved currents and proving their
conservation to all--loop orders.
While the $D(n+1,n)$ and $B(n,n)$ systems are genuine $N=1$ supersymmetric
theories,
the $A(n,n)$ models possess a global $N=2$ supersymmetry. In the
conformal case, we show that the $A(n,n)$ stress--energy tensor,
uniquely determined by the holomorphicity condition, has vanishing central
charge and it corresponds to the
stress--energy tensor of the associated topological theory.

\vglue 1.0cm}
\begin{center}
{\it Invited talk at the International Workshop}
\end{center}
{\it ``String theory, quantum gravity and the unification of the fundamental
interactions''}
\begin{center}
{\it Roma -- September 1992}
\end{center}
\vfill
\begin{flushleft}
IFUM 435/FT \\
November 1992 \\
\end{flushleft}
\end{titlepage}

\newpage
{\bf\noindent 1. Introduction \hfil}
\vglue 0.4cm
\baselineskip=14pt

In the recent past Toda field theories have been the subject of extensive
investigations in
connection with the study of $W$--algebra representations, perturbed
conformal field theories, 2d gravity and its $W$--generalizations.
The bosonic Toda systems are
constructed from a Lie algebra $\cal{G}$ and depending on whether
the algebra is affine or not the resulting theories are massive or
conformally invariant, respectively. They possess an infinite number of
conservation laws which are not spoiled by
quantum anomalies \cite{b3,b4,b5}. In the massive case
this implies the existence of factorized,
elastic S-matrices \cite{b6}, which have been determined exactly for all
unitary
theories based on affine simply--laced \cite{b7} as well as nonsimply--laced
algebras \cite{b8}.

In the region of imaginary coupling constant affine Toda
theories represent suitable perturbations of $\cal{G}$--minimal
models along a direction which maintains the
integrability properties.
The spectrum contains soliton configurations \cite{b28} and the
corresponding S--matrix has been proposed
for some of the relevant theories \cite{b20}.

It is interesting to study the fermionic generalization of these models.
Fermions can be added \cite{b1} consistently
while maintaining the basic property of classical integrability: in this case
the field theory is based on a Lie superalgebra \cite{b2}.
In general the introduction of fermions does not lead to
systems  invariant under supersymmetry transformations; indeed all
unitary fermionic Toda theories exhibit explicitly broken supersymmetry
\cite{b1}.
In order to obtain a supersymmetric model one has to consider those
superalgebras
which admit a purely fermionic system of simple roots. By selecting the ones
for which the untwisted
affine perturbation can be implemented in a supersymmetric way one is
led to consider Toda theories based on the $A(n,n)$, $ D(n+1,n)$ and $B(n,n)$
superalgebras in addition to the exceptional superalgebras $D(2,1;\a)$.
The $A(n,n)$, $ D(n+1,n)$ and $B(n,n)$ models have been recently studied in
Ref. \cite{b21,b22}.
It has been shown that both in the conformal
and in the massive cases there exist quantum higher-spin conserved currents.
Therefore the S-matrix of the massive theories is purely elastic
and factorizes into two-body processes.
At the lagrangian level these theories are nonunitary,
nonetheless the particle and soliton mass spectrum is real and the
explicit construction of exact S-matrices could be attempted.
The $A(n,n)$, $ D(n+1,n)$ and $B(n,n)$ models possess a
{\em local} $N=1$ supersymmetry at the conformal point,
so they are good candidates to describe $N=1$
superconformal minimal models. This supersymmetry becomes {\em rigid} when
the systems are perturbed by the affine interaction.
In addition the $A(n,n)$ Toda theory possesses a second
{\em rigid} supersymmetry \cite{b23}.
Its improved, holomorphic stress--energy tensor
has zero central charge and it is
the one associated to the topological version of the theory \cite{b24}.
Then it is suggestive to interpret the $A(n,n)$ Toda theories as
perturbations of the $N=2$ superconformal $A(n,n-1)$ models along
integrable, ``topological'' trajectories where the $N=2$
superconformal invariance is broken down to an $N=1$ superconformal invariance.

In this paper we focus our attention on two explicit examples, i.e. the
$B(1,1)$ and $A(1,1)$ Toda field theories. We describe the construction of
quantum higher--spin conserved currents both in the conformal and massive
cases. In the $A(1,1)$ case we discuss in detail  the role
of the $N=2$ supersymmetry in connection with topological aspects.

\vglue 0.6cm
{\bf\noindent 2. Review of the bosonic Toda theories \hfil}
\vglue 0.4cm
The general expression of the Minkowski space lagrangian for a Toda
theory based on a finite Lie algebra $\cal{G}$ is
\EQ
S = \frac{1}{\b^2}
\int d^2x \left[ \frac12 \pa_{\nu}\vec{\phi} \cdot \pa^{\nu}\vec{\phi}-
\sum_{i=1}^{r} e^{\vec{\a}_i \cdot \vec{\phi}} \right]
\label{1}
\EN
where $\vec{\a}_i$, $i=1, \cdots,r$ ($r$ = rank of $\cal{G}$) are
$r$--dimensional euclidean vectors which realize the positive simple roots
of the algebra, $\phi_a$, $a=1, \cdots,r$ are
real scalar fields and $\b$ is the coupling constant.
The classical equations of motion can be written as
\EQ
\Box \phi_j = -\sum_{i=1}^r C_{ji} e^{\phi_i}
\EN
where we have defined $\phi_j \equiv \vec{\a}_j \cdot \vec{\phi}$ and
$C_{ij} \equiv \vec{\a}_i \cdot \vec{\a}_j$.
Being the symmetric matrix $C_{ij}$ invertible
the potential in eq.(\ref{1}) does not possess minima at finite values of
the fields, so that the theory has no classical stable point.
The theory is conformally invariant since its stress--energy tensor,
suitably improved, is traceless. In order to show this explicitly it is
convenient to use light--cone coordinates:
\bea
z \equiv x^+ = \frac{1}{\sqrt{2}} (x^0 + x^1) \qquad \qquad
\bar{z}\equiv x^-= \frac{1}{\sqrt{2}}(x^0 - x^1)  \nonumber \\
\partial \equiv \partial_z = \frac{1}{\sqrt{2}} (\partial_0 + \partial_1)
\qquad \qquad \bar{\partial} \equiv \partial_{\bar{z}} = \frac{1}{\sqrt{2}}
(\partial_0 - \partial_1) \qquad \qquad \Box = 2 \partial \bar{ \partial}
\label{3}
\ena
and define light--cone components of the stress--energy tensor
\bea
&& T \equiv T_{++} = T_{00} +T_{11}+2T_{01} \nonumber \\
&& \bar{T} \equiv T_{--} = T_{00}+T_{11}-2T_{01} \nonumber \\
&& \Theta \equiv T_{+-} = T_{00}-T_{11}
\ena
Using the equations of motion from eq.(\ref{1}) it follows that
\EQ
T = -\pa \vec{\phi} \cdot \pa \vec{\phi} + \frac{2}{\b} \vec{\rho}
\cdot \pa^2 \vec{\phi}
\EN
is conserved and traceless, $\bar{\pa}T=0$, whenever $\vec{\rho}$ is the
Weyl vector of the algebra:
\EQ
\vec{\rho} = \sum_{i=1}^r \vec{\omega}_i \qquad \vec{\omega}_i \cdot
\vec{\a}_j = \delta_{ij}
\EN

To stabilize the vacuum of the theory one can add new
exponential terms to the potential in eq.(\ref{1}). In general this will
introduce a mass scale and destroy the conformal invariance of the theory.
However certain classes of perturbations do maintain the integrability
property which the theory possesses at its conformal point. This is
achieved by adding the perturbing term
\EQ
\frac{1}{\b^2} e^{\vec{\a}_0 \cdot \vec{\phi}}
\EN
where $\vec{\a}_0$ is an extra euclidean $r$--dimensional vector such that the
set $(\vec{\a}_0, \vec{\a}_1, \cdots , \vec{\a}_r)$ forms an extended,
but admissible root system
described by an affine Dynkin diagram of the algebra. The simplest
possibility is to consider the set which realizes the untwisted affine
extension of the algebra, for which $\vec{\a}_0$ is the negative of the
{\em highest} root of $\cal{G}$.
In the affine version the Toda potential has a minimum at finite values of
the fields.
One can shift the ${\phi}_i$'s so that the minimum is at $\phi_i=0$ and
the potential can be written as
\EQ
V(\vec{\phi}) = \sum_{i=0}^r q_i e^{\vec{\a}_i \cdot \vec{\phi}}
\label{2}
\EN
where $q_i$ are the Kac labels of $\cal{G}$ defined by the condition
$\sum_{i=0}^r q_i\vec{\a}_i =0$, $q_0=1$.

The classical integrability of such theories, i.e. the existence of an
infinite number of conserved currents, follows from the identification
of the Toda field equations with the integrability condition of a linear
problem. Given the $r$ generators $\vec{h}$
of the Cartan subalgebra of $\cal{G}$ and the basis vectors $e_j^{\pm}$
associated to the positive and negative roots $(\vec{\a}_0, \cdots
,\vec{\a}_r)$ satisfying the commutation relations
\EQ
[\vec{h},\vec{h}] = 0 \qquad [\vec{h},e_j^{\pm}] = \pm \vec{\a}_j e_j^{\pm}
\qquad
[e_i^{+},e_j^{-}] = \delta_{ij} \vec{h} \cdot \vec{\a}_i
\label{ccr}
\EN
one defines the Lie algebra--valued gauge connection
\EQ
A =  \pa \vec{\Phi} \cdot \vec{h} + \lambda \sum_{j=0}^r e_j^{+}
\qquad \qquad \quad
\bar{A} = \frac{1}{2\l} \sum_{j=0}^r q_j e^{\vec{\a}_j \cdot \vec{\Phi}}
e_j^{-}
\EN
where $\l$ is the spectral parameter.
The integrability condition of the linear Lax system
\EQ
(\pa +A) \chi = 0 \qquad (\bar{\pa}-\bar{A})\chi = 0
\label{Lax}
\EN
is then
\EQ
\bar{\pa} A + \pa \bar{A} + [A,\bar{A}] = 0
\label{50}
\EN
Using the commutation relations in eq.(\ref{ccr}), it is straightforward
to check that the zero curvature condition in eq.(\ref{50})
is equivalent to the Toda equations of motion with the
perturbed potential in eq.(\ref{2}).
The system (\ref{Lax}) guarantees the existence of an infinite set of
classical conserved currents \cite{b11}.
(The quantum version of these currents has been studied in Refs.
\cite{b3,b4,b5}.)

\vglue 0.6cm
{\bf \noindent 3. N=1 supersymmetric Toda theories \hfil}
\vglue 0.4cm

Toda theories that contain fermionic as well as bosonic fields are
constructed from a Lie superalgebra ${\cal G}$.
Lie superalgebras have been classified by Kac \cite{b2}
on the basis of the associated
Cartan matrix. A distinguishing feature
is given by the fact that the system of simple roots for a given Lie
superalgebra is not unique (up to a Weyl transformation). In general it is
possible to define several unequivalent
sets which have a different content of fermionic roots:
they are related by generalized Weyl transformations associated to
fermionic roots \cite{b9}.

As for the bosonic case, the classical field equations
can be viewed as the zero curvature condition on a Lax connection in $N=1$
superspace $(z, \bar{z}, \theta, \bar{\theta})$, where $z$, $\bar{z}$
are the Minkowski space light--cone coordinates defined in eq.(\ref{3})
and $\theta$, $\bar{\theta}$ are the (1,1) spinor coordinates.
The corresponding covariant derivatives
\EQ
D = \partial_{\theta} + i \theta \partial  \qquad \qquad
\bar{D} = \partial_{\bar{\theta}} - i \bar{\theta} \bar{\partial}
\EN
satisfy the commutation relations $\{D, \bar{D}\} = 0$, $D^2 =
i \partial$, $\bar{D}^2 = -i \bar{\partial}$.  We denote by
$\vec{\Phi}$ the superfields
\EQ
\Phi^a = \phi^a + \frac{1}{\sqrt{2}} \theta \psi^a + \frac{1}{\sqrt{2}}
\bar{\theta} \bar{\psi}^a + \theta \bar{\theta} F^a  \qquad \quad a=1,
\dots ,r
\EN
where $r$ is the rank of $\cal{G}$.
Introducing the $r$ generators $\vec{H}$ of the Cartan subalgebra of
$\cal{G}$, and the {\em even } ({\em odd}) generators $E_i^{\pm}$ associated
to the positive and negative {\em bosonic} ({\em fermionic}) simple roots,
satisfying the graded commutation relations
\EQ
[\vec{H},\vec{H}] = 0 \qquad [\vec{H},E_j^{\pm}] = \pm \vec{\a}_j E_j^{\pm}
\qquad
[E_i^{+},E_j^{-}\} = \delta_{ij} \vec{H} \cdot \vec{\a}_i
\label{C1}
\EN
we define the two Lie superalgebra--valued gauge superfields \cite{b1,b10}
\bea
&& U =  D \vec{\Phi} \cdot \vec{H} - \lambda \left(\sum_{j \in \cal{F}} E_j^{+}
+ \theta \sum_{j \in \cal{B}} E_j^+ \right)  \nonumber \\
&& \bar{U} = -\frac{1}{2\l} \left( \sum_{j \in \cal{F}}  q_j e^{\vec{\a}_j
\cdot
\vec{\Phi}} E_j^{-} + \bar{\theta} \sum_{j \in \cal{B}}  q_j e^{\vec{\a}_j
\cdot \vec{\Phi}} E_j^{-} \right)
\label{4}
\ena
where $\cal{B}$ and $\cal{F}$ are the two subsets of
bosonic and fermionic simple roots respectively and $\l$ is the spectral
parameter.
The integrability of the Lax pair
\EQ
(D+U)\chi = 0 \qquad (\bar{D} +\bar{U})\chi = 0
\label{150}
\EN
corresponds to the flatness condition on the Lax connection
\EQ
\bar{D} U + D \bar{U} + \{U,\bar{U}\} = 0
\label{flat}
\EN
Once again, using the commutation relations in eq.(\ref{C1}), one can show
that equation (\ref{flat}) coincides with the equations of
motion derivable from the Toda action
\EQ
S = \frac{1}{\b^2} \int d^2 z d^2 \theta \left[ D\vec{\Phi} \cdot \bar{D}
\vec{\Phi} + \sum_{i \in \cal{F}} q_i e^{\vec{\a}_i \cdot \vec{\Phi}}
+\theta \bar{\theta} \sum_{i \in \cal{B}} q_i e^{\vec{\a}_i \cdot \vec{\Phi}}
\right]
\label{101}
\EN
where  $d^2 z = dz d \bar{z}$ , $d^2 \theta = \bar{D} D$
and $\b$ is the coupling constant. If bosonic roots are present in the
Dynkin diagram of $\cal{G}$ supersymmetry is explicitly broken by the
last term in the action.

At the classical level these theories are integrable as
a consequence of the identification of the Toda field equations with the
linear problem in eq.(\ref{150}).

In order to obtain Toda theories
with $N=1$ unbroken supersymmetry one has to consider Toda models
associated to Lie superalgebras which admit a purely fermionic root system.
In the following we will restrict ourselves to these cases with the
additional requirement that the untwisted affine extension be realized by
$\vec{\a}_0 = -\sum_{i=1}^r q_i\vec{\a}_i$ with $\vec{\a}_0$ fermionic.
This selects the following superalgebras:

\noindent
1) The unitary series $A(n,n) \equiv sl(n+1,n+1;\cal{C})$,
$n \geq 1$.
The superalgebra $A(n,n)$ is not simple; it possesses a
one--dimensional graded invariant subalgebra generated by the basis element
${\bf I}_{2n+2}$. $A(n,n)$ is then defined as $sl(n+1,n+1;\cal{C})$
factored by this subspace.

\noindent
2) The orthosymplectic series $D(n+1,n) \equiv Osp(2n+2,2n;\cal{C})$,
$n \geq 1$ and
$B(n,n)=Osp(2n+1,2n;\cal{C})$, $n \geq 1$. These superalgebras are simple.

\noindent
3) Among the exceptional superalgebras only $D(2,1;\a)$, $\a \neq 0,-1$,
have a fermionic Dynkin
diagram with the affine extension also fermionic. They are deformations of the
$D(2,1)$ superalgebra.

The corresponding purely fermionic Dynkin diagrams and their
untwisted extensions are given in Fig. $1$.

The quantum conservation laws for the $A(n,n)$, $D(n+1,n)$ and
$B(n,n)$ supersymmetric Toda theories have been
studied extensively in Refs. \cite{b21,b22}. In what follows we will give the
results for the two particularly simple cases of the $B(1,1)$ and $A(1,1)$
models.

\vglue 0.6cm
{\bf \noindent 4. The ${\bf B(1,1)}$ Toda theory \hfil}
\vglue 0.4cm
We consider the $N=1$ supersymmetric Toda theory associated to the Lie
superalgebra $B(1,1) \equiv Osp(3,2)$. The fermionic Dynkin diagram
contains two simple roots ( see Fig.$1$ ). They can be realized in terms of
two--dimensional euclidean vectors:
\EQ
\vec{\a}_1 = (1,-i) \qquad \vec{\a}_2 = (0,i)
\EN
Therefore the corresponding Toda theory contains two superfields whose
dynamic is described by the following action:
\EQ
S = \frac{1}{\b^2} \int d^2 z d^2 \theta \left[ D\Phi_1 \bar{D} \Phi_1 + D
\Phi_2 \bar{D} \Phi_2 + e^{\Phi_1-i\Phi_2} + 2e^{i\Phi_2} \right]
\label{15}
\EN
The equations of motion for $\Phi_1$ and $\Phi_2$ are
\bea
D\bar{D} \Phi_1 &=& \frac12 e^{\Phi_1-i\Phi_2} \nonumber \\
D\bar{D} \Phi_2 &=& \frac12 \left(-ie^{\Phi_1-i\Phi_2} +2ie^{i\Phi_2}
\right)
\label{eq}
\ena
As described in the previous section, this theory is classically
integrable.
The most practical way to obtain explicit expressions for the
conserved supercurrents is via the Miura operator \cite{b19,b22}
\EQ
\Delta(\vec{\Phi}) =
(D+D\vec{\Phi} \cdot \vec{\l}_5)(D+D\vec{\Phi} \cdot \vec{\l}_4) \cdots
(D+D\vec{\Phi} \cdot \vec{\l}_1) = D^5 + \sum_{i=0}^4 W^{(2-\frac{i}{2})}
D^i
\label{Miura1}
\EN
where $\vec{\l}_j$, $j=1, \cdots,5$ are the weights of the fundamental
representation, $\vec{\l}_1 =-\vec{\l}_5
=\vec{\a}_1+\vec{\a}_2$, $\vec{\l}_2 =-\vec{\l}_4=\vec{\a}_2$ and $\vec{\l}_3
=0$.
Therefore eq.(\ref{Miura1}) can be written as
\EQ
(D - D\Phi_1)(D -iD\Phi_2)D(D + iD\Phi_2)(D + D\Phi_1) =
W^{(2)}+ W^{(\frac{3}{2})}D +W^{(1)} D^2 +D^5
\EN
where
\bea
W^{(1)} &=&
-i D \Phi_1 \pa \Phi_1 -i D \Phi_2 \pa \Phi_2 + D\pa \Phi_2 +2i D \pa \Phi_1
\nonumber \\
W^{(\frac{3}{2})} &=&
- i \pa^2 \Phi_2 +i D \Phi_1 D\pa \Phi_1 - (\pa \Phi_2)^2 -i D \Phi_2
D\pa \Phi_2 \nonumber \\
{}~~~~&~& +2D \Phi_1 D\pa \Phi_2  -2iD \Phi_1 D \Phi_2  \pa \Phi_2
\nonumber \\
W^{(2)} &=&
-D \pa^2\Phi_1 +D\Phi_1 \pa^2\Phi_1-i D\Phi_1 \pa^2\Phi_2 +
iD \pa \Phi_2 \pa\Phi_1
-(\pa \Phi_2)^2  D \Phi_1 \nonumber \\
{}~~~~&~&+ \pa \Phi_2 D \Phi_2 \pa \Phi_1 -i D \Phi_2
D \pa \Phi_2 D \Phi_1 \nonumber \\
&=& \frac{1}{2} D(W^{(\frac{3}{2})} +DW^{(1)})
\label{21}
\ena
Using the equations of motion (\ref{eq}) it is easy to see that the Miura
operator commutes with the spinor derivative
$\bar{D}$ so that it generates a set of superholomorphic conserved currents:
\EQ
\bar{D}W^{(s)} = 0 \qquad ~~~s=1,\frac{3}{2},2
\label{151}
\EN
We note that $W^{(1)}$ can be written as
\EQ
W^{(1)} = -iD\vec{\Phi} \cdot \pa \vec{\Phi} + 2i \vec{\rho} \cdot
D \pa \vec{\Phi}
\label{155}
\EN
where $\vec{\rho} =(1,-\frac{i}{2})$ is the Weyl vector. It is proportional
to the improved super stress--energy tensor of the theory.
Due to the conservation law (\ref{151}) the theory has a manifest $N=1$
superconformal invariance. The components of $W^{(1)}$
\bea
\left. W^{(1)}\right|_{\theta=0} &=& -\frac{i}{\sqrt{2}} \psi_1 \pa \phi_1
-\frac{i}{\sqrt{2}} \psi_2 \pa \phi_2 +i\sqrt{2} \pa \psi_1 +
\frac{1}{\sqrt{2}} \pa \psi_2 \nonumber \\
\left. DW^{(1)}\right|_{\theta=0} &=&
\pa \phi_1 \pa \phi_1 + \pa \phi_2 \pa \phi_2 +
\frac{i}{2} \psi_1 \pa \psi_1 + \frac{i}{2} \psi_2 \pa \psi_2 -2\pa^2
\phi_1 +i\pa^2 \phi_2
\ena
are respectively the generators of supersymmetry and conformal transformations.

The quantum integrability of the theory has been addressed
in Refs. \cite{b21,b22} by studying the renormalization and conservation of
the supercurrents.
There it has been shown that the holomorphic currents
in eq.(\ref{21}) maintain their form at the quantum level,
albeit the coefficients of the various terms acquire a coupling constant
dependence. We report here a brief summary of the procedure and refer for
details of the calculation to Ref.\cite{b21}.

In order to determine the quantum corrections to the classical currents
it is convenient to use massless perturbation theory which is
best suited for an all--loop analysis \cite{b5}.
The massless superspace propagators are given by
\EQ
\left \langle
\Phi_i(Z,\bar{Z}) \Phi_j(0,0)\right \rangle =
- \d_{ij} \frac {\b^2}{4\pi} \bar{D} D[log(2z \bar{z})
\d^{(2)}(\theta)]
\EN
with $(Z,\bar{Z}) \equiv (z,\theta,\bar{z},\bar{\theta})$.
The quantum lagrangian is defined by normal ordering the exponentials
in eq.(\ref{15}) ( no ultraviolet divergences ), which are then treated
as interaction terms.
The quantum conservation laws defined as
\EQ
\bar{D}_Z \left \langle W^{(s)}(Z,\bar{Z}) \right \rangle \equiv \bar{D}_Z
\left \langle W^{(s)}(Z,\bar{Z})
\exp \left(\frac{i}{\b^2}
\int d^2w d^2 \theta' {\cal L}_{int} \right) \right \rangle _0 = 0
\label{9}
\EN
could be spoiled by local anomalies
from Wick contractions of the current $W^{(s)}$ with the first order
expansion in ${\cal L}_{int}$. In fact, once the
$D$--algebra has been performed we are left with terms of the form
\EQ
\bar{D}_Z \int d^2w {\cal A}(z,\theta, \bar{z},\bar{\theta}) \bar{D}_Z
\frac{1}{(z-w)^n} {\cal B} (w,\theta,\bar{w},\bar{\theta})
\EN
where ${\cal A}$ and ${\cal B}$ are products of superfields and their
derivatives. Now, using
\EQ
\bar{D}_Z \bar{D}_Z \frac{1}{(z-w)^n}= -i \bar{\pa}_z \frac{1}{(z-w)^n}
=\frac{2\pi}{(n-1)!} \pa^{n-1}_w \d^{(2)}(z-w) \label{int}
\EN
we produce local contributions which affect the last equality in
eq.(\ref{9}). In Ref. \cite{b21} it has been proved that these potential
anomalies can be cancelled by modifying the classical
$W^{(s)}$-currents, i.e. by adding coupling constant dependent terms.
Since the calculation can be done to all--loop orders,
the exact quantum version of the conserved supercurrents is thus obtained.
For the $B(1,1)$ case one has
\EQ
W^{(1)}= -i D\Phi_1 \pa \Phi_1- i D\Phi_2 \pa \Phi_2  +(1- \frac{\b^2}{4 \pi})
D\pa \Phi_2 + 2i(1- \frac{\b^2}{8 \pi})D \pa \Phi_1
\label{17}
\EN
\bea
W^{(\frac{3}{2})}&=& -i (1-\frac{\b^2}{4\pi}) \pa^2 \Phi_2
+i(1-\frac{\b^2 }{2\pi}) D \Phi_1 D \pa \Phi_1 - (\pa \Phi_2)^2
-i D \Phi_2 D \pa \Phi_2 \nonumber \\
&~&+2(1-\frac{\b^2}{4\pi}) D\Phi_1 D \pa \Phi_2 -2i D \Phi_1 D \Phi_2
\pa \Phi_2
\label{18}
\ena
The  quantum spin--$2$ current can be obtained from
eqs.(\ref{17}) and (\ref{18})
since  $W^{(2)}$ is linearly dependent on  $W^{(1)}$ and $W^{(\frac{3}{2})}$.
Eq.(\ref{17}) gives the quantum expression of the super stress--energy
tensor. The corresponding central charge is
\EQ
c = 3 + 24 \vec{\rho}_q^{~2}
\EN
where the ``quantum'' Weyl vector is
$\vec{\rho}_q = \left( 1-\frac{\b^2}{8\pi}, -\frac{i}{2} +i\frac{\b^2}{8\pi}
\right)$.

A massive perturbation of the $B(1,1)$ Toda theory is realized by adding to
the action (\ref{15}) an exponential term associated to the lowest fermionic
root
\EQ
\vec{\a}_0 = -(q_1\vec{\a}_1 + q_2 \vec{\a}_2) = (-1,-i)
\EN
The action for the $B^{(1)}(1,1)$ Toda theory is
\EQ
S = \frac{1}{\b^2} \int d^2z d^2 \theta \left[ D\vec{\Phi} \cdot \bar{D}
\vec{\Phi} + e^{\Phi_1-i\Phi_2} + 2e^{i\Phi_2} + e^{-\Phi_1-i\Phi_2}
\right]
\label{30}
\EN
The classical mass spectrum can be obtained from the second order expansion of
the lagrangian:
\EQ
{\cal L}^{(2)} \equiv  -M_1\Phi_1^2 -M_2\Phi_2^2 = \Phi_1^2-2\Phi_2^2
\EN
It follows that the bosonic masses are
\EQ
m_1^2 = 2M_1^2 = 2 \quad ; \quad m_2^2 = 2M_2^2 = 8
\EN
We note that the classical mass spectrum is real despite the
manifest non--unitarity of the Lagrangian.

Due to the precence of a mass scale the perturbed theory is no longer
conformally invariant. The super stress--energy tensor acquires a
non--vanishing trace and its conservation law takes the form
\EQ
\bar{D} J^{(s)} + D \bar{J}^{(s)} = 0
\EN
with $s=1$.
The study of higher--spin ($s>1$)conserved currents reveals
that the affine theory
does not have conserved quantities at levels $s=\frac{3}{2}$ and $s=2$.
At the classical level the first non--trivial conserved current has spin
$3$ and it can be expressed in terms of the $W$--currents as
\EQ
J^{(3)} = W^{(1)}DW^{(1)} + 2W^{(1)}W^{(\frac{3}{2})}
\EN
While in the conformal case the two terms are separately holomorphic
currents, in the affine case only the linear combination above satisfies
the classical conservation law.

At the quantum level a lengthy calculation leads to a quantum spin--$3$
conserved current \cite{b21} and to a corresponding charge
\EQ
Q^{(3)} = \int dz d\theta J^{(3)}(z,\bar{z},\theta,\bar{\theta})
\EN
which satisfies $\bar{D} Q^{(3)} = 0$.

The existence of higher--spin conserved currents insures the quantum
integrability of the system and the exact, factorizable
S--matrix could be determined. The actual construction of an S--matrix is
complicated by the presence of solitons in the spectrum, due to the
periodicity of the potential under the shift
\EQ
\Phi_2 \rightarrow \Phi_2 + 2\pi k
\EN
$k$ being any integer. In particular, soliton solutions are given by
$\Phi_1 = 0$ and $\Phi_2$ assuming the field configurations of the super
sine--Gordon solitons \cite{b17}. The solitonic masses are {\em real}.

The previous analysis can be extended to the $B(n,n)$ theories with $n>1$.
These are $N=1$ superconformal models with central charge $c= \frac{3}{2}r
+ 24\vec{\rho}^{~2}$, where $r$ is the rank of the algebra and $\vec{\rho}$
its Weyl vector. Massive perturbations associated to the lowest fermionic
root give rise to nonunitary integrable systems (the action is not hermitian).
However, the bosonic mass spectrum is real
\bea
m_k^2 &=& 8 {\sin}^2 \frac{2\pi k}{h}
\qquad k=1, \cdots ,2n-1 \nonumber \\
m_{2n}^2 &=& 2
\ena
For theories with $n>1$ the classical mass ratios are affected by quantum
corrections \cite{b22}.

These theories contain solitons. Indeed, setting to zero the fields which
appear in the potential with real couplings the generic $B^{(1)}(n,n)$
theory reduces to $n$ decoupled sine--Gordon systems. Solitons have a real
mass spectrum. The reality of the elementary particle and soliton masses
leads to the expectation that, even if unitarity is not manifest at the
lagrangian level, these theories admit a unitary restriction.

\vglue 0.6cm
{\bf \noindent 5. The ${\bf A(1,1)}$ Toda theory \hfil}
\vglue 0.4cm
{\it\noindent 5.1. Conservation laws}
\vglue 0.1cm
We consider now the supersymmetric Toda theory associated to the $A(1,1)$
superalgebra. This algebra is defined as $sl(2,2;{\cal C})$ factored by
the one--dimensional invariant subspace generated by ${\bf I}_4$.
The purely fermionic Dynkin diagram and its untwisted fermionic extension
are given in Fig.$1$. Since this superalgebra is not simple, the number
of simple roots exceeds the rank by one.
Their explicit expressions can be given in terms of two--dimensional
vectors as
\bea
\vec{\a}_1 &=& = (1,-i) \nonumber \\
\vec\a_2 &=& = (1,i) \nonumber \\
\vec{\a}_3 &=& -\vec{ \a}_1
\label{root3}
\ena
The corresponding supersymmetric action is
\EQ
S = \frac{1}{\b^2} \int d^2z d^2\theta \left[ D\Phi_1 \bar{D} \Phi_1 + D \Phi_2
\bar{D} \Phi_2 + e^{\Phi_1 -i\Phi_2} + e^{\Phi_1+i\Phi_2} +
e^{-\Phi_1+i\Phi_2} \right]
\label{19}
\EN
and the equations of motion for the two superfields are
\bea
D \bar{D} \Phi_1 &=& \frac12 \left( e^{\Phi_1 -i\Phi_2} + e^{\Phi_1+i\Phi_2} -
e^{-\Phi_1+i\Phi_2} \right) \nonumber \\
D \bar{D} \Phi_2 &=& \frac12 \left( -ie^{\Phi_1 -i\Phi_2} +
ie^{\Phi_1+i\Phi_2} + ie^{-\Phi_1+i\Phi_2} \right)
\label{154}
\ena
To establish the existence of classical conserved currents one can
construct the Miura operator (Ref. \cite{b22})
\EQ
D(D-D\Phi_1 +iD\Phi_2)(D+2iD\Phi_2)(D+D\Phi_1+iD\Phi_2)=D^4 + \sum_{i=0}^2
W^{(\frac{3}{2}-\frac{i}{2})}~D^i
\label{20}
\EN
which gives the classical currents
\bea
W^{(\frac12)} &=& -2\pa \Phi_2 + 2iD\Phi_2 D\Phi_1 \nonumber \\
W^{(1)} &=& iD\Phi_1 \pa \Phi_1 + iD\Phi_2 \pa \Phi_2 - iD\pa \Phi_1 - D\pa
\Phi_2 + i D(D\Phi_2 D\Phi_1) \nonumber \\
W^{(\frac{3}{2})} &=& -D(W^{(1)} - DW^{(\frac12)})
\label{160}
\ena
The Miura operator in eq.(\ref{20}) does
{\em not} commute with the spinor derivative $\bar{D}$, so in general the
$W$--currents are not holomorphic.
However, using the equations of motion (\ref{154}) one obtains
\EQ
\bar{D} W^{(1)}=0
\label{c1}
\EN
and
\EQ
\bar{D} W^{(\frac12)} = -2 D e^{-\Phi_1 +i\Phi_2} \qquad \qquad ~~
\bar{D} W^{(\frac{3}{2})} = -2i D\pa e^{-\Phi_1 +i\Phi_2}
\label{c2}
\EN
The spin--$1$ current is holomorphic and gives the
classical traceless super stress--energy tensor of the theory,
whereas $W^{(\frac12)}$ and
$W^{(\frac{3}{2})}$ are not holomorphic but still conserved.
The existence of a spin--$1$ holomorphic current is a general
property of all the $A(n,n)$ theories which describe then $N=1$
superconformal models.

The expressions in eq.(\ref{160}) give also the correct currents for the
quantum system since in this case no corrections appear in the study of
the conservation laws ({\ref{c1}),(\ref{c2}).

Writing $W^{(1)}$ in components we obtain the supersymmetry current
\EQ
\left. W^{(1)}\right|_{\theta=0} = \frac{i}{\sqrt{2}} \psi_1 \pa\phi_1
+\frac{i}{\sqrt{2}} \psi_2 \pa\phi_2 -\frac{i}{\sqrt{2}} \pa \psi_1
-\frac{1}{\sqrt{2}} \pa \psi_2 -\frac{1}{\sqrt{2}} \psi_1\pa\phi_2 +
\frac{1}{\sqrt{2}} \psi_2\pa\phi_1
\EN
and the stress--energy tensor which generates conformal tranformations
\EQ
\left. DW^{(1)}\right|_{\theta=0} = -\pa \vec{\phi} \cdot \pa \vec{\phi} -
\frac{i}{2} \vec{\psi} \cdot \pa \vec{\psi} +\pa^2\phi_1 -i\pa^2\phi_2 -
\frac12 \pa(\psi_2\psi_1)
\label{156}
\EN
We note that in this case the traceless condition requires
the addition of the fermionic improvement term
$\frac12 \pa(\psi_1\psi_2)$. This term gives a negative contribution to the
central charge which turns out to be identically zero. In the next
subsection we will discuss
some of the implications of these results.

We turn now to the massive perturbation of the theory
realized by adding to the potential the exponential associated to the
fermionic lowest root
\EQ
\vec{\a}_0 = -(\vec{\a}_1+\vec{\a}_2+\vec{\a}_3) = (-1,-i)
\EN
The affine Toda action
\EQ
S = \frac{1}{\b^2} \int d^2z d^2 \theta \left[ D \vec{\Phi} \cdot \bar{D}
\vec{\Phi} + e^{\Phi_1-i\Phi_2} + e^{\Phi_1+i\Phi_2} + e^{-\Phi_1+i\Phi_2}
+ e^{-\Phi_1-i\Phi_2} \right]
\label{31}
\EN
is manifestly hermitian and symmetric under $\Phi_1 \rightarrow
-\Phi_1$ and $\Phi_2 \rightarrow -\Phi_2$. The bosonic classical masses are
\EQ
m_1^2 = m_2^2 = 8
\EN
The classical mass ratios are stable under quantum corrections \cite{b22}.

The system is quantum integrable.
In addition to the $W^{(\frac12)}$ current and the super stress--energy tensor
$W^{(1)}$ which are conserved also in the affine theory, the first
nontrivial conserved current appears again at $s=3$.
Its exact quantum expression is (for details of the calculation see Ref.
\cite{b22})
\bea
J^{(3)} &=& -i\left(1 - \frac{11\b^4}{16\pi^2}\right)
[D\pa \Phi_1 \pa^2 \Phi_1 + D\pa \Phi_2 \pa^2 \Phi_2] -i
\left(1-\frac{3\b^2}{4\pi} \right)
D\Phi_1 (\pa \Phi_1)^3 \nonumber \\
&~~& +i \left( 1 +\frac{3\b^2}{4\pi} \right) D\Phi_2 (\pa \Phi_2)^3
-i \left( 3-\frac{9\b^2}{4\pi} \right) (\pa \Phi_1)^2 D\Phi_2 \pa \Phi_2
\nonumber \\
&~~& +i\left( 3-\frac{3\b^2}{4\pi} \right) (\pa \Phi_2)^2 D\Phi_1 \pa
\Phi_1
- \frac{3\b^2}{\pi} D\Phi_1 D\pa \Phi_1 D\Phi_2 \pa \Phi_2
\ena
The spectrum of the theory contains soliton solutions which are the same as
in the super sine--Gordon model.

The $A(n,n)$ theories with $n>1$ describe $N=1$ superconformal models with
central charge $c=0$. Their untwisted fermionic affinization is described
by an action which is not hermitian, nonetheless
the bosonic particle mass spectrum is real
\EQ
m_k^2 = m_{2n+1-k}^2 = 8{\sin}^2\frac{\pi k}{n+1} \quad k=1,\cdots,n
\EN
and stable under quantum corrections.
Solitonic solutions with real masses have also been determined \cite{b23,b26}.

\vglue 0.4cm
{\it\noindent 5.1. N=2 supersymmetry}
\vglue 0.1cm

In addition to the explicit $N=1$ supersymmetry the $A(n,n)$ models possess
a second supersymmetry \cite{b23}.
We discuss here the $n=1$ case. The actions in eqs.(\ref{19}), (\ref{31})
for the conformal and massive $A(1,1)$ theories respectively,
are invariant under $N=2$ supersymmetry transformations
\bea
\delta \Phi_1 &=& \varepsilon Q_1 \Phi_1 \qquad \qquad
\delta\Phi_1 = \xi Q_2 \Phi_2 \nonumber \\
\delta \Phi_2 &=& \varepsilon Q_1 \Phi_2 \qquad \qquad
\delta\Phi_2  =-\xi Q_2\Phi_1
\ena
generated by the two anticommuting operators
\EQ
Q_1 = \pa_{\theta} -i\theta \pa \qquad \qquad Q_2 \equiv D =
\pa_{\theta}+i\theta \pa
\label{157}
\EN
In terms of the superfields
\bea
\Phi_+ &= \Phi_1+i\Phi_2 \equiv \phi_+ +\frac{1}{\sqrt{2}}\theta\psi_+
+\frac{1}{\sqrt{2}} \bar{\theta} \bar{\psi}_+ + \theta \bar{\theta} F_+ \\
\Phi_- &= \Phi_1-i\Phi_2 \equiv \phi_- +\frac{1}{\sqrt{2}}\theta\psi_-
+\frac{1}{\sqrt{2}}\bar{\theta} \bar{\psi}_- + \theta \bar{\theta} F_-
\ena
we can write the general action as
\EQ
S = \frac{1}{\b^2} \int d^2z d^2\theta \left[ D\Phi_+\bar{D}\Phi_- +
e^{\Phi_-} + e^{\Phi_+} + g_1 e^{-\Phi_-} + g_2 e^{-\Phi_+} \right]
\EN
Setting $g_1=1$, $g_2=0$ it gives the $A(1,1)$ $N=1$ superconformal invariant
action in eq.(\ref{19}); for $g_1=g_2=1$ it gives the affine $A^{(1)}(1,1)$
action (\ref{31}), whereas for $g_1=g_2=0$ it reduces to the $N=2$
superconformal $A(1,0)$ theory \cite{b10}.
In components we have
\bea
S &=& \int d^2z \left[ \pa \phi_+\bar{\pa}\phi_- + \frac{i}{2}\psi_+
\bar{\pa}\psi_- -\frac{i}{2}\bar{\psi}_+\pa\bar{\psi}_- \right. \nonumber \\
&~&~~~~~~~ \left. -e^{\phi_++\phi_-} + g_1 e^{\phi_+-\phi_-} +g_2 e^{-\phi_+
+\phi_-} -g_1 g_2 e^{-\phi_+-\phi_-} \right.\nonumber \\
&~&~~~~~~~ \left. +\frac12 \bar{\psi}_+\psi_+
\left( e^{\phi_+} +g_2 e^{-\phi_+} \right)
+\frac12 \bar{\psi}_- \psi_- \left( e^{\phi_-} +g_1 e^{-\phi_-} \right) \right]
\label{200}
\ena
Correspondingly we can reexpress the $W$--currents in eq.(\ref{160}) as
\bea
W^{(\frac12)} &=& D\Phi_+ D\Phi_- + i\pa \Phi_+ -i\pa \Phi_- \nonumber \\
W^{(1)} &=& i D\Phi_- \pa \Phi_+ -iD\pa \Phi_-
\ena
They satisfy the conservation equations
\bea
\bar{D} W^{(\frac12)} &=& -2D\left( g_1 e^{-\Phi_-} - g_2 e^{-\Phi_+} \right)
\nonumber \\
\bar{D} W^{(1)} &=&  -2i\pa \left( g_2 e^{-\Phi_+} \right)
\label{158}
\ena
In particular, when $g_1=g_2=0$ they are holomorphic.

We discuss now the various symmetries generated by these currents. At the
component level we define
\bea
J &\equiv& \left. W^{(\frac12)} \right|_{\theta=0} = \frac12 \psi_+\psi_- +
i\pa\phi_+-i\pa\phi_-  \nonumber \\
Q_+ &\equiv& \sqrt{2} i \left[~ \left. W^{(1)} \right|_{\theta=0} -
\left. DW^{(\frac12)} \right|_{\theta=0} ~ \right] =
-\psi_+\pa\phi_- + \pa\psi_+ \nonumber \\
Q_- &\equiv& \sqrt{2}i \left. W^{(1)} \right|_{\theta=0} =
-\psi_-\pa\phi_+ +\pa\psi_-  \nonumber \\
T &\equiv& \left. DW^{(1)} \right|_{\theta=0} = -\pa\phi_-\pa\phi_+
-\frac{i}{2}\psi_-\pa\psi_+ +\pa^2\phi_-
\label{n2}
\ena
The $Q_{\pm}$ currents generate the two supersymmetry transformations,
$J$ is a $U(1)$ generator and $T$ is the
stress--energy tensor of the theory. As previously emphasized, the central
charge vanishes. Using the expressions of the currents in eq.(\ref{n2}) we
can construct the corresponding charges from which the infinitesimal
transformations of the fields can be easily computed. In general, from the
conservation equation
\EQ
\bar{\pa} J^{(s)} = \pa \bar{J}^{(s)}
\EN
it follows that the corresponding time independent charge,
$\frac{dQ^{(s)}}{dt}=0$ is
\EQ
Q^{(s)} = \int dz J^{(s)} + \int d\bar{z} \bar{J}^{(s)}
\label{201}
\EN
Using the equations of motion derivable from the action in eq.(\ref{200})
we have
\bea
\bar{\pa} J &=& \pa \left[ \frac12 \bar{\psi}_- \bar{\psi}_+ +i\bar{\pa}
\phi_+ -i\bar{\pa} \phi_- \right] \nonumber \\
\bar{\pa} Q_+ &=& \pa \left( -2ig_1 \bar{\psi}_- e^{-\phi_-} \right) + \pa
\left[ \bar{\pa} \psi_+ + i\bar{\psi}_- \left( e^{\phi_-} +g_1e^{-\phi_-}
\right) \right] \nonumber \\
\bar{\pa} Q_- &=& \pa \left( -2ig_2 \bar{\psi}_+ e^{-\phi_+} \right) + \pa
\left[ \bar{\pa} \psi_- + i\bar{\psi}_+ \left( e^{\phi_+} +g_2e^{-\phi_+}
\right) \right] \nonumber \\
\bar{\pa} T &=& \pa \left[ -2g_2 \left( e^{-\phi_+ +\phi_-} -g_1
e^{-\phi_+-\phi_-} \right) +g_2  \psi_+ \bar{\psi}_+ e^{-\phi_+} \right]
\nonumber \\
{}~~~~~&~~& + i \pa \left[ \psi_+ \bar{\pa} \psi_- + i \psi_+ \bar{\psi}_+
\left( e^{\phi_+} + g_2e^{-\phi_+} \right) \right] \nonumber \\
{}~~~~~&~~&
+ \frac{i}{2} \pa \left[ \psi_- \bar{\pa} \psi_+ + i \psi_- \bar{\psi}_-
\left( e^{\phi_-} + g_1 e^{-\phi_-} \right) \right] \nonumber \\
{}~~~~~&~~& + \pa \left[ \pa \bar{\pa} \phi_- + \left( e^{\phi_+} +
g_2e^{-\phi_+} \right) \left( e^{\phi_-} - g_1e^{-\phi_-} \right) \right.
\nonumber \\
{}~~~~~&~~& ~~~~~\left. + \frac12
\psi_+ \bar{\psi}_+ \left( e^{\phi_+} -g_2 e^{-\phi_+} \right) \right]
\ena
In the r.h.s. we have included terms which vanish on--shell in such a way
that the charges constructed according to eq.(\ref{201}) do generate
invariances of the action. We obtain
\\ \noindent
1) $U(1)$--transformations:
\bea
\d_J \phi_{\pm} &=& 0 \nonumber \\
\d_J \psi_{\pm} &=& \mp \z \psi_{\pm} \nonumber \\
\d_J \bar{\psi}_{\pm} &=& \pm \z \bar{\psi}_{\pm}
\label{t1}
\ena
2) $Q_+$--supersymmetry transformations:
\bea
\d_{Q_+} \phi_+ &=& \xi \psi_+ \nonumber \\
\d_{Q_+} \phi_- &=& 0 \nonumber \\
\d_{Q_+} \psi_+ &=& 0 \nonumber \\
\d_{Q_+} \psi_- &=& 2i\xi \pa\phi_- +2i\pa\xi \nonumber \\
\d_{Q_+} \bar{\psi}_+ &=& 2\xi \left( e^{\phi_-} -g_1e^{-\phi_-} \right)
\nonumber \\
\d_{Q_+} \bar{\psi}_- &=& 0
\label{t2}
\ena
3) $Q_-$--supersymmetry transformations:
\bea
\d_{Q_-} \phi_+ &=& 0 \nonumber \\
\d_{Q_-} \phi_- &=& \varepsilon \psi_- \nonumber \\
\d_{Q_-} \psi_+ &=& 2i\varepsilon \pa\phi_+ +2i\pa\varepsilon \nonumber \\
\d_{Q_-} \psi_- &=& 0 \nonumber \\
\d_{Q_-} \bar{\psi}_+ &=& 0 \nonumber \\
\d_{Q_-} \bar{\psi}_- &=& 2\varepsilon \left( e^{\phi_+} -g_2e^{-\phi_+}
\right)
\label{t3}
\ena
4) Conformal transformations generated by $T$:
\bea
\d_T\phi_+ &=& \l\pa\phi_+ +\pa \l \nonumber \\
\d_T\phi_- &=& \l\pa\phi_- \nonumber \\
\d_T\psi_+ &=& \l\pa\psi_+ \nonumber \\
\d_T\psi_- &=& \l\pa\psi_- +\pa\l\psi_- \nonumber \\
\d_T\bar{\psi}_{\pm} &=& \l \pa\bar{\psi}_{\pm}
\label{t4}
\ena

In the $A(1,0)$ theory both $W^{(\frac12)}$ and $W^{(1)}$ are superholomorphic
currents and the action is invariant under the {\em local} transformations
given above.
The holomorphic currents $J$, $Q_{\pm}$ and the ``untwisted'' stress--energy
tensor $T' \equiv T-\frac12 \pa J$ generate a $N=2$ superconformal algebra.
The topological version \cite{b27} of this $N=2$ superconformal theory is
described by
the $c=0$ stress-energy tensor $T$. The role of the BRST charge is played
by the supersymmetry charge $Q_+$ which satisfies the nihilpotency condition
$Q_+^2 =0$. It is easy to check that the stress-energy tensor can be written as
\EQ
T = \{Q_{\rm BRST}, Q_-\}
\label{159}
\EN

The $A(1,1)$ Toda system is obtained by adding
the $g_1$ perturbation. As explicitly shown in eq.(\ref{158}) the
$W^{(\frac12)}$ current is no longer holomorphic, so that the $U(1)$ and $Q_+$
symmetries become {\em
rigid}. However, the stress-energy tensor $T$ is still holomorphic
and the theory preserves a $N=1$ superconformal invariance. We note that being
$T$ the {\em only} holomorphic spin--$1$ current, the $A(1,1)$ Toda  theory
is naturally in its topological, conformal phase.

Finally we can switch the $g_2$--perturbation on and reach the
$A^{(1)}(1,1)$
massive theory. In this case neither current is holomorphic and
all the symmetries become {\em rigid}. The current $T$ is the
stress--energy tensor associated to the
topological version of the theory.

\vglue 0.6cm
{\bf \noindent 6. Conclusions \hfil}
\vglue 0.4cm
We have analyzed the quantum integrability of the $A(1,1)$ and $B(1,1)$
supersymmteric Toda theories and of their affine untwisted extensions.
We have chosen these two specific examples since they are good representatives
of the more general supersymmetric theories based on the $A(n,n)$, $D(n+1,n)$
and $B(n,n)$ superalgebras. Most of the relevant issues could be discussed in
a simple setting and easily generalized: in particular we have given an
explicit lagrangian realization in $N=1$ superspace and constructed the
quantum higher--spin conserved currents both in the conformal and massive
case.

The $B(1,1)$ Toda theory is characterized by two holomorphic conserved
supercurrents of spin $1$ and $\frac{3}{2}$. The components of $W^{(1)}$ are
the generators of supersymmetry and conformal transformations, while
$W^{(\frac{3}{2})}$ contains the generators of higher--spin transformations.

The supersymmetric theory based on the $A(1,1)$ superalgebra possesses a second
supersymmetry in addition to the one manifest in the $N=1$ superspace
formalism.
In fact the theory admits two conserved supercurrents of spin $\frac{1}{2}$
and $1$.
The spin--$1$ current is holomorphic and its components are one supersymmetry
current and the stress-energy tensor. The $W^{(\frac{1}{2})}$--current is not
holomorphic, but still conserved: at the component level it contains the
generator of $U(1)$--transformations and  the second supersymmetry current. As
emphasized in the previous section, the stress--energy tensor that we have
obtained via the Miura operator construction has zero central charge and it
can be written as a BRST--commutator.

We have briefly discussed the implications of the existence of these quantum
symmetries for the $A(1,0)$, $A(1,1)$ and $A^{(1)}(1,1)$ theories. A more
detailed analysis will be presented in a future publication \cite{b26}.

\vglue 0.6cm
{\bf \noindent 7. Acknowledgements \hfil}
\vglue 0.4cm
We wish to thank Camillo Imbimbo and Mario Pernici for valuable discussions
and suggestions.

\newpage
{\bf\noindent 8. References \hfil}
\vglue 0.4cm

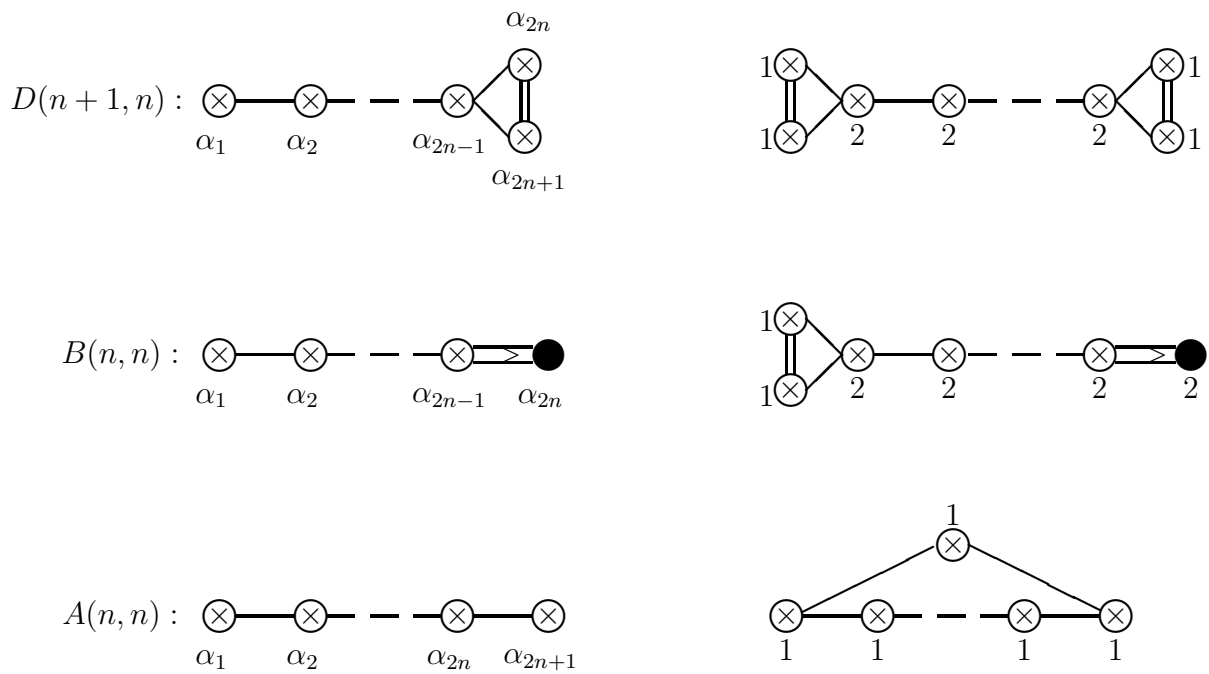
\begin{figure}


\begin{center}
\begin{tabular}{lclc}
\setlength{\unitlength}{1.5pt}
\begin{picture}(400,140)(-31,-90)
\thicklines

\put(-31,10){\makebox(0,0){$D(n+1,n):$}}
\put(0,10){\circle{8}}
\put(0,10){\makebox(0,0){$\times$}}
\put(-2,-1){\makebox(0,0){$\a_1$}}
\put(4,10){\line(1,0){15}}
\put(23,10){\circle{8}}
\put(23,10){\makebox(0,0){$\times$}}
\put(21,-1){\makebox(0,0){$\a_2$}}
\put(27,10){\line(1,0){7}}
\put(38,10){\line(1,0){7}}
\put(49,10){\line(1,0){7}}
\put(60,10){\circle{8}}
\put(60,10){\makebox(0,0){$\times$}}
\put(58,-1){\makebox(0,0){$\a_{2n-1}$}}
\put(64,10){\line(1,1){9}}
\put(64,10){\line(1,-1){9}}
\put(77,19){\circle{8}}
\put(77,19){\makebox(0,0){$\times$}}
\put(78,30){\makebox(0,0){$\a_{2n}$}}
\put(77,1){\circle{8}}
\put(77,1){\makebox(0,0){$\times$}}
\put(78,-10){\makebox(0,0){$\a_{2n+1}$}}
\put(76,5){\line(0,1){10}}
\put(78,5){\line(0,1){10}}

\put(-25,-54){\makebox(0,0){$B(n,n):$}}
\put(0,-54){\circle{8}}
\put(0,-54){\makebox(0,0){$\times$}}
\put(-2,-65){\makebox(0,0){$\a_1$}}
\put(4,-54){\line(1,0){15}}
\put(23,-54){\circle{8}}
\put(23,-54){\makebox(0,0){$\times$}}
\put(21,-65){\makebox(0,0){$\a_2$}}
\put(27,-54){\line(1,0){7}}
\put(38,-54){\line(1,0){7}}
\put(49,-54){\line(1,0){7}}
\put(60,-54){\circle{8}}
\put(60,-54){\makebox(0,0){$\times$}}
\put(58,-65){\makebox(0,0){$\a_{2n-1}$}}
\put(64,-52){\line(1,0){15}}
\put(64,-56){\line(1,0){15}}
\put(73,-54){\makebox(0,0){$>$}}
\put(83,-54){\circle*{8}}
\put(81,-65){\makebox(0,0){$\a_{2n}$}}

\put(-25,-120){\makebox(0,0){$A(n,n):$}}
\put(0,-120){\circle{8}}
\put(0,-120){\makebox(0,0){$\times$}}
\put(-2,-131){\makebox(0,0){$\a_1$}}
\put(4,-120){\line(1,0){15}}
\put(23,-120){\circle{8}}
\put(23,-120){\makebox(0,0){$\times$}}
\put(21,-131){\makebox(0,0){$\a_2$}}
\put(27,-120){\line(1,0){7}}
\put(38,-120){\line(1,0){7}}
\put(49,-120){\line(1,0){7}}
\put(60,-120){\circle{8}}
\put(60,-120){\makebox(0,0){$\times$}}
\put(58,-131){\makebox(0,0){$\a_{2n}$}}
\put(64,-120){\line(1,0){15}}
\put(83,-120){\circle{8}}
\put(83,-120){\makebox(0,0){$\times$}}
\put(81,-131){\makebox(0,0){$\a_{2n+1}$}}

\put(148,19){\line(1,-1){9}}
\put(148,1){\line(1,1){9}}
\put(144,19){\circle{8}}
\put(144,19){\makebox(0,0){$\times$}}
\put(136,16){\makebox{$1$}}
\put(144,1){\circle{8}}
\put(144,1){\makebox(0,0){$\times$}}
\put(136,-2){\makebox{$1$}}
\put(143,5){\line(0,1){10}}
\put(145,5){\line(0,1){10}}
\put(161,10){\circle{8}}
\put(161,10){\makebox(0,0){$\times$}}
\put(159,-1){\makebox{$2$}}
\put(165,10){\line(1,0){15}}
\put(184,10){\circle{8}}
\put(184,10){\makebox(0,0){$\times$}}
\put(182,-1){\makebox{$2$}}
\put(189,10){\line(1,0){7}}
\put(200,10){\line(1,0){7}}
\put(211,10){\line(1,0){7}}
\put(222,10){\circle{8}}
\put(222,10){\makebox(0,0){$\times$}}
\put(220,-1){\makebox{$2$}}
\put(226,10){\line(1,1){9}}
\put(226,10){\line(1,-1){9}}
\put(239,19){\circle{8}}
\put(239,19){\makebox(0,0){$\times$}}
\put(244,16){\makebox{$1$}}
\put(239,1){\circle{8}}
\put(239,1){\makebox(0,0){$\times$}}
\put(244,-2){\makebox{$1$}}
\put(238,5){\line(0,1){10}}
\put(240,5){\line(0,1){10}}

\put(148,-45){\line(1,-1){9}}
\put(148,-63){\line(1,1){9}}
\put(144,-45){\circle{8}}
\put(144,-45){\makebox(0,0){$\times$}}
\put(136,-48){\makebox{$1$}}
\put(144,-63){\circle{8}}
\put(144,-63){\makebox(0,0){$\times$}}
\put(136,-67){\makebox{$1$}}
\put(143,-59){\line(0,1){10}}
\put(145,-59){\line(0,1){10}}
\put(161,-54){\circle{8}}
\put(161,-54){\makebox(0,0){$\times$}}
\put(159,-65){\makebox{$2$}}
\put(165,-54){\line(1,0){15}}
\put(184,-54){\circle{8}}
\put(184,-54){\makebox(0,0){$\times$}}
\put(182,-65){\makebox{$2$}}
\put(189,-54){\line(1,0){7}}
\put(200,-54){\line(1,0){7}}
\put(211,-54){\line(1,0){7}}
\put(222,-54){\circle{8}}
\put(222,-54){\makebox(0,0){$\times$}}
\put(220,-65){\makebox{$2$}}
\put(226,-52){\line(1,0){15}}
\put(226,-56){\line(1,0){15}}
\put(233,-56){\makebox{$>$}}
\put(245,-54){\circle*{8}}
\put(243,-65){\makebox{$2$}}

\put(143,-120){\circle{8}}
\put(143,-120){\makebox(0,0){$\times$}}
\put(141,-131){\makebox{$1$}}
\put(147,-120){\line(1,0){15}}
\put(166,-120){\circle{8}}
\put(166,-120){\makebox(0,0){$\times$}}
\put(164,-131){\makebox{$1$}}
\put(170,-120){\line(1,0){7}}
\put(181,-120){\line(1,0){7}}
\put(192,-120){\line(1,0){7}}
\put(203,-120){\circle{8}}
\put(203,-120){\makebox(0,0){$\times$}}
\put(201,-131){\makebox{$1$}}
\put(207,-120){\line(1,0){15}}
\put(226,-120){\circle{8}}
\put(226,-120){\makebox(0,0){$\times$}}
\put(224,-131){\makebox{$1$}}
\put(147,-119){\line(2,1){33}}
\put(189,-102){\line(2,-1){33}}
\put(185,-102){\circle{8}}
\put(185,-102){\makebox(0,0){$\times$}}
\put(183,-97){\makebox{$1$}}

\end{picture}
\end{tabular}
\end{center}
\vskip 2.5truecm
\caption {{\em Fermionic Dynkin diagrams for Lie superalgebras and their affine
extensions. The crossed circles denote fermionic roots
with vanishing norm.
In the affine diagrams the Kac labels are explicitly shown.}}
\end{figure}

\end{document}